\documentclass{jfm}

\usepackage{amssymb,amsmath}
\usepackage{epsfig}
\usepackage{bm}
\usepackage{color}
\usepackage{epstopdf}
\usepackage{graphicx}
\usepackage{natbib}

\newcommand\beq{\begin{equation}}
\newcommand\eeq{\end{equation}}
\newcommand\beqa{\begin{eqnarray}}
\newcommand\eeqa{\end{eqnarray}}

\title[Towards a better understanding of granular flows]{Towards a better understanding of granular flows}

\author[Vicente Garz\'o]
{V\ls I\ls C\ls E\ls N\ls T\ls E\ls \ns G\ls A\ls R\ls Z\ls \'O\ls
$\dagger$}
\affiliation{Departamento de F\'isica, Instituto Universitario de Computaci\'on Cient\'ifica Avanzada (ICCAEx), Universidad de Extremadura, Avda. de Elvas s/n, 06006  Badajoz (Spain)}

\begin{document}

\maketitle

\begin{abstract}

Understanding the transport of particles immersed in a carrier fluid (bedload transport) is still an exciting challenge. Among the different types of gas-solid flows, when the dynamics of solid particles is essentially dominated by collisions between them, kinetic theory can be considered as a reliable tool to derive continuum approaches from a fundamental point of view. In a recent paper, \cite{ChChB23} have proposed a two-fluid model based on modifications to a classical kinetic theory model \cite[]{GD99a}. First, in contrast to the classical model, the model proposed by Chassagne \emph{et al.} takes into account the interparticle friction not only in the radial distribution function but also through an effective restitution coefficient in the rate of dissipation term of granular temperature. As a second modification, at the top of the bed where the volume fraction is quite small, the model accounts for the saltation regime in the continuum framework. The theoretical results derived from the model agree with discrete simulations for moderate and high densities and they are also consistent with experiments. Thus, the model proposed by \cite{ChChB23} helps to a better understanding on the combined impact of friction and inelasticity on the macroscopic properties of granular flows.


\end{abstract}

$\dagger${Email address for correspondence: vicenteg@unex.es}

\date{\today}
\maketitle

\section{Introduction}
\label{sec1}

It is well known that granular materials are ubiquitous in nature and play an important role in many industrial processes, such as those involving pharmaceutical, agricultural, or construction products among others. The fact that these materials are widely used in industry is likely one of the main reasons explaining why their understanding has attracted the attention of physicists and engineers in the past few years. Apart from their practical interest, their study is interesting by itself since they behave differently depending on the external conditions to which they are subjected. For instance, grains contained in a jar will behave as a solid, liquid, or gas depending on how they are shaken. Even when the motion of grains is quite similar to the random motion of atoms or molecules of a molecular fluid (rapid flow conditions), granular flows differ from conventional fluid flows since the size of grains is macroscopic and hence, their collisions are inelastic.

In the dilute regime, rapid flow conditions can be achieved when the energy dissipated by collisions is compensated for by the energy supplied to the system by external excitations. In this regime, kinetic theory (which can be considered as a mesoscopic description intermediate between statistical mechanics and hydrodynamics), conveniently adapted to account for dissipative dynamics, has been employed in the past few years as the starting point to derive hydrodynamic equations with explicit forms of the transport coefficients. Usually, a simple model has been considered in the granular literature: a gas of \emph{smooth} hard spheres where the inelasticity in collisions is accounted for via a (positive) constant coefficient of normal restitution $e\leq1$. The \emph{inelastic} versions of the classical Boltzmann and Enskog kinetic equations for dilute and moderate dense gases, respectively, have been proposed to determine the dynamic properties of granular gases \cite[]{G19}. However, technical difficulties in the analysis of the above equations entailed approximations that limited the applicability of the first attempts \cite[]{LSJCh84,JR85a} to weakly dissipative granular flows ($e\lesssim 1$). Years later, \cite{GD99a} (GD-theory) solved the Enskog equation by means of the Chapman--Enskog method and obtained expressions for the Navier--Stokes transport coefficients for the whole range of values of the coefficient of restitution $e$.

Although grains in nature are usually surrounded by a fluid like water or air, all the above models \cite[]{LSJCh84,JR85a,GD99a} neglect the effect of the interstitial fluid on the dynamics of grains. Needless to say, understanding the flow of solid particles immersed in one or more fluid phases is a very intricate problem. Among the different types of multiphase flows, a particularly interesting set corresponds to the so-called particle-laden suspensions,
in which small, immiscible, and typically dilute particles are immersed in a carrier fluid (for instance, fine aerosol particles in air) \cite[]{S20}. When the dynamics of grains is essentially dominated by collisions among them, kinetic theory is again a convenient tool to describe this type of flows. Due to complexity of the problem, a coarse-grained approach is usually adopted, where the effect of the background fluid on grains is accounted for through an effective fluid-solid force. In some situations \cite[]{TK95}, only a viscous drag force (proportional to the instantaneous particle velocity) is considered, while other more realistic models \cite[]{GTSH12,GGG19a} also incorporate a stochastic Langevin-like term defined in terms of the background temperature $T_\text{b}$.

On the other hand, for dense granular flows, kinetic theory has been so far incapable of describing this type of regime. These type of flows are generally described with the so-called $\mu(I)$ rheology \cite[]{FP08}, where a phenomenological law (obtained by fitting experimental and discrete simulation data) relates the stress rate $\mu$ with the inertial number $I$. Although this phenomenological theory gives good predictions in the dense regime, it fails for dilute systems.

A recent paper by \cite{ChChB23} proposes a frictional-collisional two-fluid model based on kinetic theory for modeling bedload transport. The proposed model (which employs the GD frictionless kinetic theory as a baseline) covers dilute and dense regimes and exhibits a good agreement with DEM simulations and experiments \cite[]{NC18}.

\begin{figure}
\begin{center}
\resizebox{4.5cm}{!}{\includegraphics{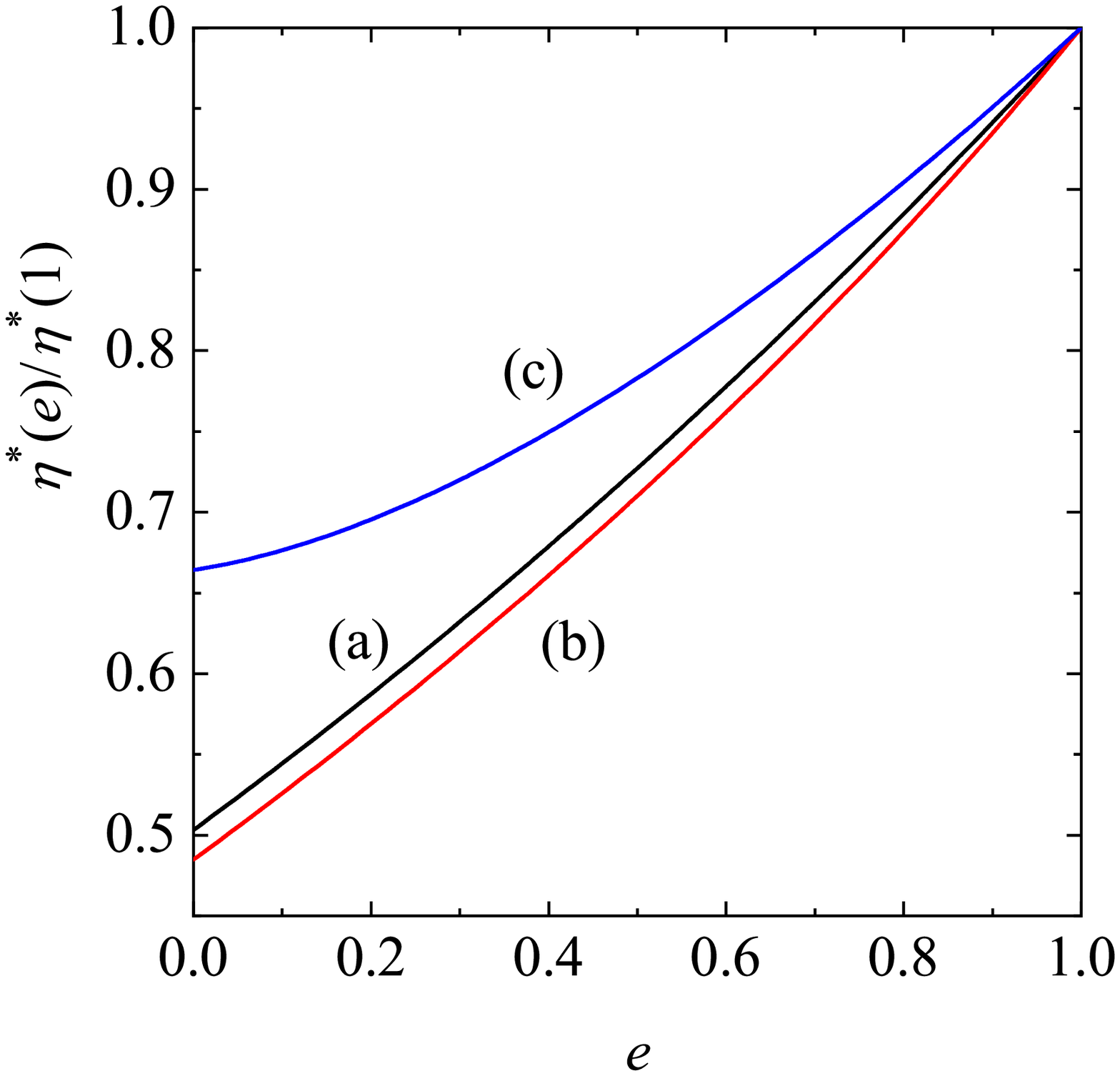}}\resizebox{4.5cm}{!}
{\includegraphics{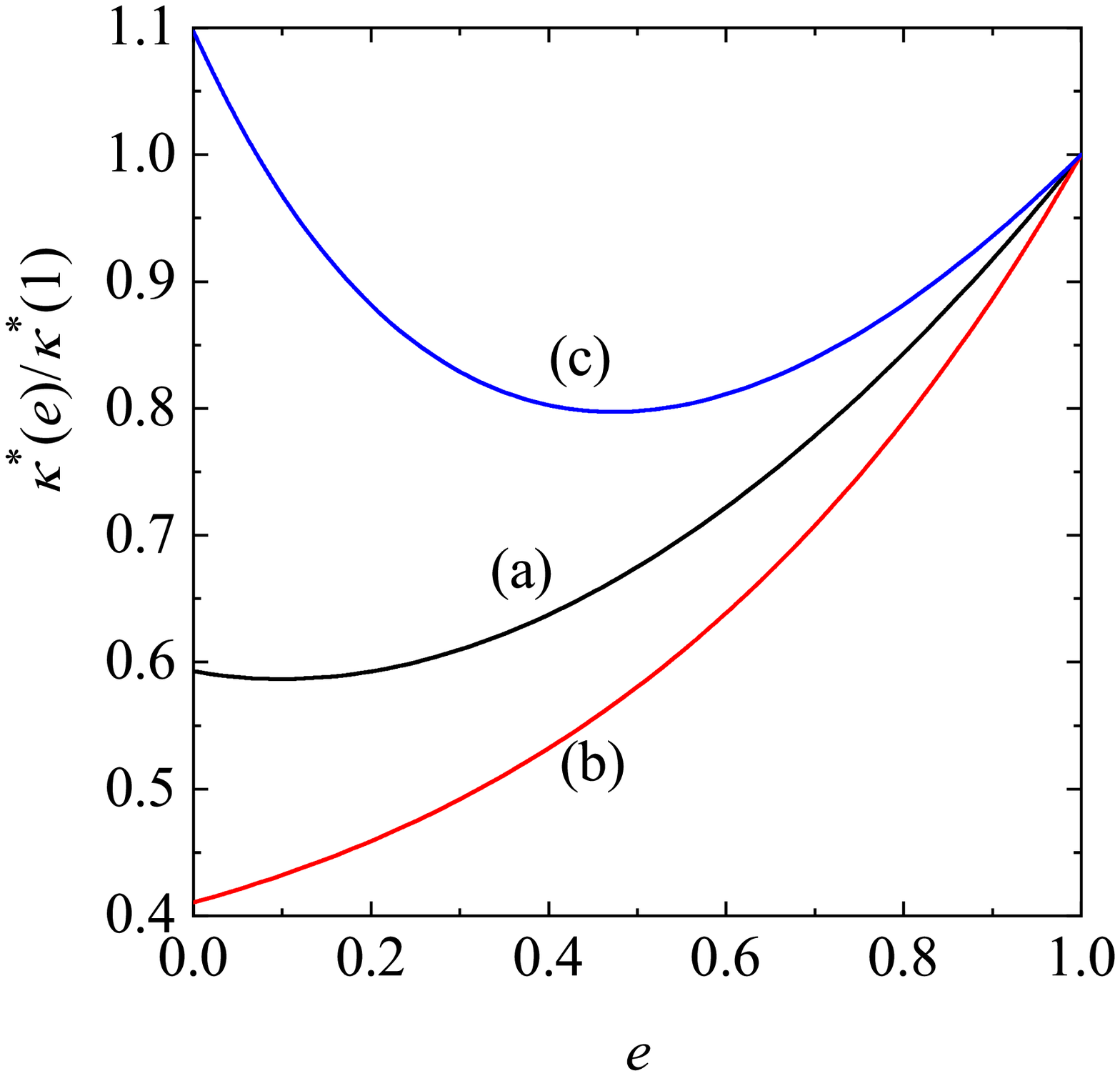}}\resizebox{4.5cm}{!}{\includegraphics{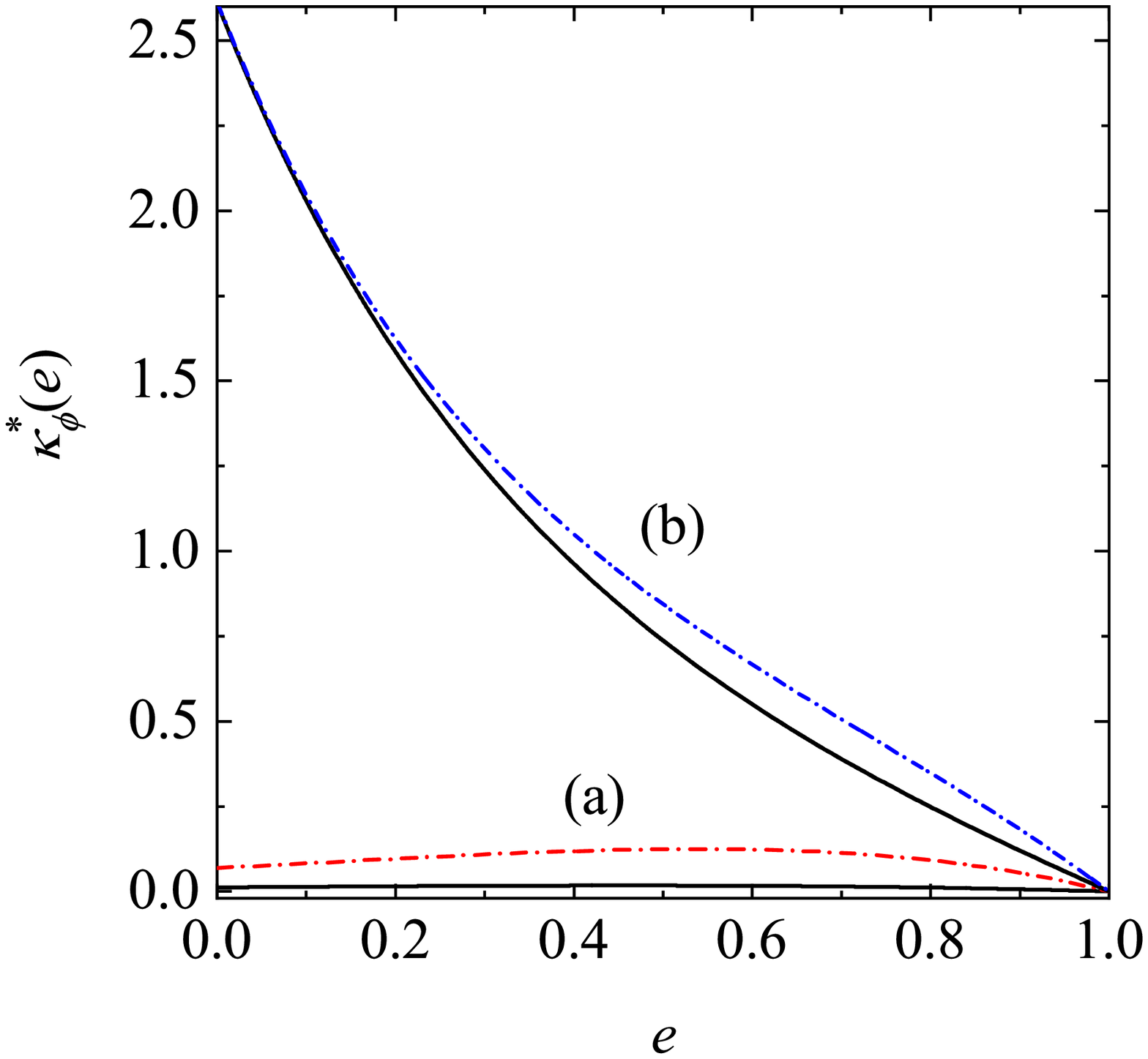}}
\end{center}
\vspace{0.1cm}
\caption{Plot of the (reduced) coefficients $\eta^*(e)/\eta^*(1)$, $\kappa^*(e)/\kappa^*(1)$, and $\kappa_\phi^*(e)$ versus the coefficient of restitution $e$ for $\mu^p=0.8$ and $T_\text{b}^*=T_\text{b}/m d^2 \gamma^2=1$. In the case of the coefficients $\eta^*(e)/\eta^*(1)$ and $\kappa^*(e)/\kappa^*(1)$ (left and middle panels), the lines (a), (b), and (c) correspond to the results obtained from the CBC-theory, the GG-theory, and the GD-theory, respectively, for $\phi=0.2$. In the case of the coefficient  $\kappa_\phi^*$ (right panel), the solid (dash-dotted) lines refer to the results obtained from the GG-theory (a) and the GD-theory (b) for $\phi=0.1$ ($\phi=0.2$).
\label{fig1}}
\end{figure}

\section{Overview}
\label{sec2}

To mitigate the discrepancies observed between the theoretical predictions of the frictionless GD-theory with those obtained in DEM simulations, \cite{ChChB23} (CBC-theory) corrects the above kinetic theory in the following ways. First, since the GD-theory neglects the effect of the surrounding fluid, a drag force term between the fluid and particle phases is introduced. This drag force leads to a dissipation term in the balance equation for the granular temperature due to fluctuating motion particle. Second, given that simulations clearly show that interparticle friction adds another source of energy dissipation different from that of inelasticity, an \emph{effective} restitution coefficient $e_\text{eff}$ \cite[]{JZ02} is introduced in the expression of the rate of dissipation of granular temperature. The effective coefficient $e_\text{eff}(\mu^p)$ is smaller than the (constant) coefficient of restitution $e$ and depends on the friction coefficient $\mu^p$. In addition, since interparticle friction also affects the geometrical packing structure of the granular flow, they modify the usual form of the radial distribution function $g_0$ (which takes into account spatial correlations) by adding a $\mu^p$-dependent term adjusted to reproduce the discrete simulations. A third modification to the GD-theory refers to the saltation regime observed at the top of the bed, where the volume fraction is small. Since this regime is essentially dominated by the fluid drag force, it cannot be reproduced by the GD-theory. Although in most of the previous works saltation is treated as a boundary condition, \cite{ChChB23} model saltation in a continuum framework. While the saltation contribution $\eta^{\text{salt}}$ to the shear viscosity $\eta$ was obtained in previous works \cite[]{JCV10}, \cite{ChChB23} propose that the saltation contribution $\kappa^{\text{salt}}$ to the thermal conductivity $\kappa$ is simply given by $\kappa^{\text{salt}}=\eta^{\text{salt}}/\sigma$, where $\sigma=0.5$ is an empirical constant found by fitting to the DEM simulations.

To put the results in a proper context, it is interesting to compare the predictions for the Navier--Stokes transport coefficients obtained from the GD-theory, the CBC-theory and the GG-theory \cite[]{GGG19a}. While the former theory is for a \emph{dry} (no fluid phase) granular gas, the two latter take into account the impact of the fluid phase on grains. To perform a clean comparison between the CBC and GG theories, the fluid drag force coefficient $K$ appearing in \cite{ChChB23} may be related to the Stokes drift coefficient $\gamma$ of the GG-theory by the relation $\gamma=(1-\phi)K/\rho^p$, where $\phi$ is the solid volume fraction, $\rho^p=m/(\frac{\pi}{6}d^3)$ is the particle density, $m$ is the particle mass, and $d$ is the diameter of the sphere. In terms of the granular temperature $T$, the transport coefficients can be written as $\eta=d\rho^p \sqrt{T/m}\eta^*$, $\kappa=d\rho^p \sqrt{T/m^3}\kappa^*$, and $\kappa_\phi=T^{3/2}/(d^2 \phi\sqrt{m})\kappa_\phi^*$. Here, $\kappa_\phi$ is the transport coefficient linking the heat flux with the solid volume fraction gradient. This contribution (which vanishes for elastic collisions) to the heat flux was neglected by \cite{ChChB23}. Figure \ref{fig1} shows the dependence of the dimensionless coefficients $\eta^*(e)/\eta^*(1)$, $\kappa^*(e)/\kappa^*(1)$ and $\kappa_\phi^*(e)$ on $e$, as given by the three theories. For this moderate value of the solid volume fraction ($\phi=0.2$), we observe that the predictions of the (frictionless) GG-theory for the coefficients $\eta^*$ and $\kappa^*$ agree well (especially in the case of the shear viscosity) with those obtained from the CBC-theory, even for quite strong inelasticities ($e\gtrsim 0.5$). This is likely due to the fact that, while the effect of the friction coefficient $\mu^p$ on $g_0$ is relatively small, the impact of the fluid phase on transport is quite important. This latter aspect is not considered in the GD-theory and hence, it exhibits differences with the CBC-theory which are much more significant than those found with the GG-theory. Another interesting conclusion of the GG-theory is the negligible impact of the coefficient $\kappa_\phi^*$ on the heat transport (see the right panel of Fig.\ \ref{fig1}); this result clearly differs from the one obtained in the GD-model (where $\kappa_\phi^*$ can be even larger than the thermal conductivity coefficient $\kappa^*$ in a frictionless dry granular fluid) but agrees with the simulation data of \cite{ChChB23}.

One of the weaknesses of kinetic theory is its inability to predict dense granular flows. However, the results derived from the CBC-theory (based on the combination of kinetic theory with a frictional model) reproduces the $\mu(I)$ rheology in the dense regime. Although recent attempts to model dense granular flows with kinetic theory have been based on the introduction of a correlation length in the dissipation term, the results obtained by \cite{ChChB23} suggest that dense granular flows are essentially dominated by the competition between elastic-frictional and kinetic-collisional stresses rather than the development of velocity correlations (breakdown of molecular chaos hypothesis) for high densities.

Finally, a comparison with experiments \cite[]{NC18} of the two-fluid model proposed by \cite{ChChB23}, when turbulence is negligible, has also shown a good agreement. In this sense, the present model can be taken as a starting point to adapt it to the study of more complex configurations.

\section{Future}
\label{sec3}

The paper by \cite{ChChB23} shows the potentiality of kinetic theory to accurately describe bedload transport. Their results suggest that a complicated extension of the GD frictionless kinetic theory to account for the combined effect of the coefficients of friction and normal restitution on the transport coefficients may not be necessary to capture the behavior of granular flows for dilute and dense regimes. As a potential extension of the CBC-theory to account for the effect of interparticle friction on transport in a more rigorous way one could consider a granular gas of inelastic rough hard spheres, where a constant coefficient of tangential restitution characterizes the ratio between the magnitude of the tangential component of the relative velocity after and before collision. In this context, as a first step, one should extend the results derived by \cite{KSG14} to moderate densities and consider this theory as a baseline to account for the saltation regime in the continuum framework, as made by \cite{ChChB23}.

\vspace{0.25cm}
\noindent \textbf{Fundings.} Financial support from Grant PID2020-112936GB-I00 and from Grants IB20079 funded by Junta de Extremadura (Spain) and by ERDF A way of making Europe is acknowledged.\\

\noindent\textbf{Declaration of interests.} The author reports no conflict of interest.\\

\noindent\textbf{Author ORCIDs.}
\\Vicente Garz\'o: https://orcid.org/0000-0001-6531-9328



\begin{thebibliography}{14}
\expandafter\ifx\csname natexlab\endcsname\relax\def\natexlab#1{#1}\fi
\def\au#1{#1} \def\ed#1{#1} \def\yr#1{#1}\def\at#1{#1}\def\jt#1{\textit{#1}}
  \def\bt#1{#1}\def\bvol#1{\textbf{#1}} \def\vol#1{#1} \def\pg#1{#1}
  \def\publ#1{#1}\def\arxiv#1{#1}\def\org#1{#1}\def\st#1{\textit{#1}}

\bibitem[Chassagne {\em et~al.\/}(2023)Chassagne, Bonamy \& Chauchat]{ChChB23}
{\sc \au{Chassagne, R.}, \au{Bonamy, C.} \& \au{Chauchat, J.}} \yr{2023}  \at{A
  frictional-collisional model for bedload transport based on kinetic theory of
  granular flows: discrete and continuum approaches}.  \jt{J. Fluid Mech.}
  \bvol{964},  \pg{A27}.

\bibitem[Forterre \& Pouliquen(2008)]{FP08}
{\sc \au{Forterre, Y.} \& \au{Pouliquen, O.}} \yr{2008}  \at{Flows of dense
  granular media}.  \jt{Annu. Rev. Fluid Mech.}  \bvol{40},  \pg{1--24}.

\bibitem[Garz\'o(2019)]{G19}
{\sc \au{Garz\'o, V.}} \yr{2019} {\em Granular Gaseous Flows\/}.
  \publ{Springer Nature, Cham}.

\bibitem[Garz\'o \& Dufty(1999)]{GD99a}
{\sc \au{Garz\'o, V.} \& \au{Dufty, J.~W.}} \yr{1999}  \at{Dense fluid
  transport for inelastic hard spheres}.  \jt{Phys. Rev. E}  \bvol{59},
  \pg{5895--5911}.

\bibitem[Garz\'o {\em et~al.\/}(2012)Garz\'o, Tenneti, Subramaniam \&
  Hrenya]{GTSH12}
{\sc \au{Garz\'o, V.}, \au{Tenneti, S.}, \au{Subramaniam, S.} \& \au{Hrenya,
  C.~M.}} \yr{2012}  \at{Enskog kinetic theory for monodisperse gas-solid
  flows}.  \jt{J. Fluid Mech.}  \bvol{712},  \pg{129--168}.

\bibitem[G\'omez~Gonz\'alez \& Garz\'o(2019)]{GGG19a}
{\sc \au{G\'omez~Gonz\'alez, R.} \& \au{Garz\'o, V.}} \yr{2019}  \at{Transport
  coefficients for granular suspensions at moderate densities}.  \jt{J. Stat.
  Mech.}  \bvol{093204}.

\bibitem[Jenkins {\em et~al.\/}(2010)Jenkins, Cantat \& Valance]{JCV10}
{\sc \au{Jenkins, J.~T.}, \au{Cantat, I.} \& \au{Valance, A.}} \yr{2010}
  \at{Continuum model for steady, fully developed saltation above a horizontal
  particle bed}.  \jt{Phys. Rev. E}  \bvol{82},  \pg{020301(R)}.

\bibitem[Jenkins \& Richman(1985)]{JR85a}
{\sc \au{Jenkins, J.~T.} \& \au{Richman, M.~W.}} \yr{1985}  \at{Kinetic theory
  for plane flows of a dense gas of identical, rough, inelastic, circular
  disks}.  \jt{Phys. Fluids}  \bvol{28},  \pg{3485--3493}.

\bibitem[Jenkins \& Zhang(2002)]{JZ02}
{\sc \au{Jenkins, J.~T.} \& \au{Zhang, C.}} \yr{2002}  \at{Kinetic theory for
  identical, frictional, nearly elastic spheres}.  \jt{Phys. Fluids}
  \bvol{14},  \pg{1228--1235}.

\bibitem[Kremer {\em et~al.\/}(2014)Kremer, Santos \& Garz\'o]{KSG14}
{\sc \au{Kremer, G.~M.}, \au{Santos, A.} \& \au{Garz\'o, V.}} \yr{2014}
  \at{Transport coefficients of a granular gas of inelastic rough hard
  spheres}.  \jt{Phys. Rev. E}  \bvol{90},  \pg{022205}.

\bibitem[Lun {\em et~al.\/}(1984)Lun, Savage, Jeffrey \& Chepurniy]{LSJCh84}
{\sc \au{Lun, C. K.~K.}, \au{Savage, S.~B.}, \au{Jeffrey, D.~J.} \&
  \au{Chepurniy, N.}} \yr{1984}  \at{Kinetic theories for granular flow:
  inelastic particles in {C}ouette flow and slightly inelastic particles in a
  general flowfield}.  \jt{J. Fluid Mech.}  \bvol{140},  \pg{223--256}.

\bibitem[Ni \& Capart(2018)]{NC18}
{\sc \au{Ni, W.-J.} \& \au{Capart, H.}} \yr{2018}  \at{Stresses and drag in
  turbulent bed load from refractive index-matched experiments}.  \jt{Geophys.
  Res. Lett.}  \bvol{45},  \pg{7000--7009}.

\bibitem[Subramaniam(2020)]{S20}
{\sc \au{Subramaniam, S.}} \yr{2020}  \at{Multiphase flows: {R}ich physics,
  challenging theory, and big simulations}.  \jt{Phys. Rev. Fluids}  \bvol{5},
  \pg{110520}.

\bibitem[Tsao \& Koch(1995)]{TK95}
{\sc \au{Tsao, H-K} \& \au{Koch, D.~L.}} \yr{1995}  \at{Simple shear flows of
  dilute gas--solid suspensions}.  \jt{J. Fluid Mech.}  \bvol{296},
  \pg{211--245}.

\end{thebibliography}


\end{document}